\documentclass[pra,twocolumn]{revtex4}
\usepackage{graphicx}
\usepackage{amssymb}
\usepackage{amsmath}
\usepackage{amsbsy}
\usepackage{color,soul}
\usepackage{upgreek}
\usepackage{siunitx}
\usepackage{tikz}
\usepackage{soul}

\usepackage{multirow}

\usepackage{hyperref}

\hypersetup{
    colorlinks=true,
    linkcolor=blue,
    filecolor=magenta,      
    urlcolor=blue,
    citecolor=red
}
 
\begin{document}

\author{\bf E. Giglio}
\author{\bf M. L\'eger}

\title{Stabilizing the ion beam transmission through tapered glass capillaries}
\date{\today}
\affiliation{Centre de Recherche sur les Ions, les Mat\'eriaux et la Photonique (CIMAP), Normandie Univ, ENSICAEN, UNICAEN, CEA, CNRS, F-14000 Caen, France}


\begin{abstract}
When an ion beam is injected into a tapered insulating capillary, the induced  self-organized radial Coulomb potential in the capillary is able to focus the beam like an electrostatic lens. However, because of the  continued accumulation of charge in the capillary, an equilibrium is rarely attained and the injected beam is eventually "Coulomb" blocked by the capillary's potential.
We propose an original add-on to the capillary setup, which can be tested experimentally and which is expected to hinder the Coulomb blocking. We investigate numerically the benefits and limits of the modified capillary setup. We show in how far the intensity and emittance of the injected beam control the transmission rate through conically tapered capillaries. Our results indicate that the add-on succeeds to stabilize the asymptotic transmission rate for a larger range of beam intensities and emittances, while reaching a near optimal transmitted fraction up to 90\%.
\end{abstract}

\maketitle

\section{Introduction}

When an ion beam is injected into a tapered  
glass capillary of macroscopic dimensions, most of the injected charge accumulates in the capillary wall and only a small  fraction of the ion beam is transmitted. 
If the beam axis is aligned with the symmetry axis of the capillary, the accumulated charge is expected to have axial symmetry, which generates an axisymmetric  electric potential in the capillary. With a growing amount of accumulated charge, the self-organized electric field in the capillary focuses an increasingly larger  fraction of the injected ion beam through the capillary outlet. The transmitted current density  $j_\text{out}$ can be  orders of magnitude higher than the current density $j_\text{in}$ of the injected beam. 
We call this mechanism the self-organized  radial focusing of an ion beam by an insulating tapered capillary, and it was recently investigated by several authors \cite{Schweigler_2011,Chen,Giglio_PRA_2018,Maurya}.

Self-organized radial focusing by insulating macro-capillaries is usually observed for ion beams in the keV energy range. 
In a pioneering work, Schweigler \textit{et  al.} simulated the trajectories of  8 keV Ar$^{8+}$ ions  through tapered capillary  of different shapes and outlet diameters. 
The compression factor $f_\text{c}=j_\text{out}/j_\text{in }$ they deduced from the simulations was found typically between 2 and 5, depending on the charge depletion rate used in the simulation \cite{Schweigler_2011}. Using a 14 mm short glass capillary, Chen \textit{et al.} \cite{Chen} succeeded to focus a 90 keV  O$^{6+}$ beam and found a compression factor $f_\text{c}$ of about 4.5. 

Recently, it was shown, in a combined experimental and theoretical work,  that a tapered glass capillary can focus efficiently a 2.3 keV Ar$^+$ ion beam if the self-organized potential inside the capillary is able to reach 70 \% of the extraction potential of the ion source \cite{Giglio_PRA_2018}. The authors found a maximum value of $f_\text{c}$ of about 11 and showed that the latter depends on the intensity of the injected beam. They  also showed that the focused beam, after hours of transmission, eventually faded away. Indeed, with the capillary setup presented in \cite{Giglio_PRA_2018}, the blocking of the transmission of the focused beam could not be avoided because the potential in the capillary reached inevitably the extraction potential of the  ion source. We refer to the latter as \textsl{Coulomb blocking}.

Similarly, charge dissipation and self focusing limit in high current density ion beam transport through tapered micro glass capillaries was investigated experimentally in \cite{Maurya}. Using an Ar beam with an extraction bias voltage of 10 keV, the limit on capillary outlet diameter for focusing of high current density was studied. The authors reported a compression factor $f_\text{c}$ varying from 2 to 700 for an outlet diameter varying from 800 to 20 $\mu$m. A limit in the transmission is however observed where the self focusing of the beam ceases, as the inner diameter of the capillary is reduced.

{\color{blue}
Two other mechanisms, able to induce a compression of the  transmitted current density, were identified in the past. The first one is due to "transverse" compression by injected charge patches and appears typically when the capillary is not aligned with the beam axis.
The second one, is dominant for ion energies in the MeV range and is due to surface scattering. As they compete with the radial self-organized focusing, we will discuss them here briefly.
 }

With their funnel shaped glass capillary, Ikeda \textit{et al.} \cite{Ikeda} reported a compression factor of about 10 for the transmitted Ar$^{8+}$ beam extracted by a bias voltage of 1 kV. However, the  compression observed in \cite{Ikeda} was most probably not due to radial focusing, but was rather induced by  multiple reflections of the beam by multiple injected charge patches. Indeed, a factor $f_c=10$ was observed after only 2 pC were injected, which is way too low to induced a potential of 700 V (70\% of the extraction potential), necessary for radial-focusing. 
Beam compression by patches in capillaries was demonstrated using simulations in  \cite{Stolterfoht13-1} and \cite{Giglio_PRA_2018}, with the latter including a detailed discussion of the involved mechanism. {\color{blue} Experimental evidence of transverse compression by tapered capillaries tilted with respect to the beam axis was given in \cite{Kreller_2011,Gruber2014,Stoltherfoht_PRA_2015}. By varying the  angle $\beta$ between the capillary and the beam axis, a maximum of  $f_\text{c}=3$  was found for $\beta=0.3\deg$ in \cite{Gruber2014,Stoltherfoht_PRA_2015} and a maximum of $f_\text{c}=8$ for $\beta=0.75\deg$ in \cite{Kreller_2011}. In both cases, no enhancement was found for zero tilt angle, excluding  
thus the possibility that the enhancement was due to axially symmetric self-organized potential.

For ions extracted by bias potentials of 100 kV and more, self-organized radial focusing was  not yet observed.
The transmission of a 100 keV proton beam through micro-sized conical capillaries was recently investigated numerically in \cite{Yanga_EPJD_2020}. The authors found that the compression factor $f_\text{c}$ increased with decreasing taper angle up to a value of 3. They noted however that the beam compression came mainly from small angle scattering rather than charge induced radial focusing. We assume that the potential inside the capillary never reached the required 70 kV necessary for efficient radial focusing.  
Similarly, Hespeels \textit{et al.} concluded  that the compression factor of 2  observed for a 1.7 MeV H$^+$ beam was based
on the scattering of  ions by the inner
surface \cite{Hespeels_2015}, indicating that self-organized focusing may be limited to ions extracted by a potential well below the MV.
For a recent review on micro-beams production of ions with MeV energy, see  \cite{Ikeda_review_QUBS_2020}. 
Note that small angle scattering is always present, also for ions in the keV range, but the scattered ions pick up an electron from the surface, resulting in neutral atoms that are no longer guided. On position sensitive detectors, those neutral atoms produce typically  a more or less neutral  ring, as shown for example in \cite{Giglio_PRA_2018}.
}

In this work, the authors tackle the issues reported in \cite{Giglio_PRA_2018} and to some extent in \cite{Maurya}, that is, the Coulomb blocking of the transmitted focused beam. 
We discuss briefly the radial focusing mechanism in tapered glass capillaries and show in how far the emittance and intensity of the injected ion beam influence the rate and stability of the transmitted beam. Then, we present an add-on to the capillary setup that can be implemented experimentally without larger difficulty and that should avoid the Coulomb blocking. We  test with our home-made numerical code \textsf{InCa4D} the pertinence and efficiency of the proposed solution. The code is CPU efficient (typically over $10^7$ trajectories per day on a single modern CPU core), which allowed us to investigate the asymptotic behavior of the transmission rate.

%

%

\section{Modeling}

\subsection{Capillary setups}

\begin{figure}[ht!]
\begin{center}
\includegraphics[scale=0.5]{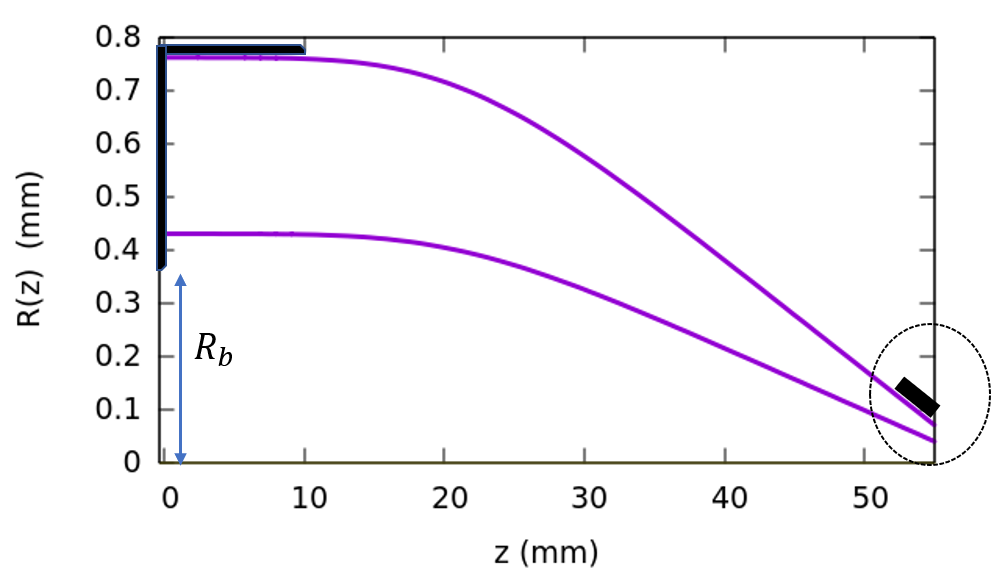}
\end{center}
\caption{Profile of the  inner  and outer glass-vacuum interfaces  of the capillary. Black solid lines stand for the grounded conducting electrodes. $R_b=0.4$ mm is the radius of the hole in the grounded collimator plate in front of the capillary. Setup (INS) and (GND) differ by the 3 mm short grounded electrode that covers the tip (circled) and which is absent in setup (INS).\label{fig_profile}}
\end{figure}

The tip of the glass capillary in \cite{Giglio_PRA_2018} was slightly funnel-shaped, which we do not expect to be optimal for stabilizing the transmission rate of a focused beam. 
For our numerical study, we chose a tapered capillary with a conical tip. 
The axisymmetric tapered capillary used in our simulation has a length of $h=55$ mm. The inner radii at the entrance and exit of the capillary are respectively $R_\text{0}=430$ $\upmu$m and $R_h=50$ $\upmu$m. The  tapered profile of the capillary is given, in cylindrical coordinates, by the radial function,
\begin{equation}
R(z)=R_h + (R_0-R_h)\,\text{erf}\left[\left(\frac{h-z}{\Delta z}\right)^n \right]^{1/n}  , \; 0\le z \le h \label{eq_shape}
\end{equation}
where $\Delta z=37$ controls the length of the tapered part of the capillary and $n=2.5$ the stiffness (curvature) of the junction  between the beginning straight and following tapered part. Expression (\ref{eq_shape}) allows us to generate different capillary profiles, from straight to conical capillaries, mainly by modifying the parameters $\Delta z$, $R_h$ and $n$. 
The ratio between the outer and inner radii of the capillary is assumed constant for all $z$ and equal to 1.77. The latter value corresponds to the ratio of the inner and outer diameters of the capillary at the entrance, and is motivated by our observation that the said ratio is conserved during the pulling process of a heated glass tube. The profiles of the outer and inner surface of the capillary are shown in Fig.\ref{fig_profile}. The conical tip of the inner surface is characterized by a half opening angle of $\beta= 10.8$ mrad. The entrance and the first 8 mm of the outer surface of the capillary are covered by a grounded electrode. 
The injected beam is collimated by the entrance electrode to a diameter of $2 R_b=0.8$ mm. The tapered capillary allows thus a maximal theoretical compression factor $f_\text{c}$ of $(R_b/R_h)^2 =64$. 
%
%
The chosen geometry of the tapered capillary corresponds typically to our home-made glass capillaries, we use in our experimental setup in CIMAP, Caen.  
%

In this study, we will distinguish two configurations. 
The first configuration, labeled (INS), corresponds to the case where no special treatment was applied to the capillary. The tapered part of the capillary is simply insulating glass. 
For the second configuration, labeled (GND), the last 3 millimeters of the outer surface are grounded. The grounded electrode at the tip, highlighted in Fig. \ref{fig_profile} by the dotted circle, can be realized experimentally, by painting the tip with a conducting carbon paste or silver paint and then connecting the electrode to the ground with a thin ($\varnothing < 50$ $\upmu$m) wire that has little torsion, so as not to bend the capillary tip. Grounding parts of a capillary, for studying the stability of the transmission rate in tapered capillaries, was already suggested in \cite{Ikeda_APL_2016}.

\subsection{Charge dynamics}

The simulations presented in this work have been obtained with our numerical code, \textsf{InCa4D}. {\color{blue}The theoretical model on which the code is based uses the following assumptions. Each Ar$^+$ ion that hits the inner capillary surface ejects $N$ electrons from the impact point. Of the $N$ ejected electrons, one is captured by the Ar$^+$ ion, which becomes neutralized. The other $N-1$ electrons are reabsorbed by the inner surface, presumably near the impact point. Thus, in the process, only one electron has been removed from the impact point at the surface, or alternatively, 1 hole has been injected. The injected hole is  quickly trapped by hole-centers near the interface. 
Holes and electrons in borosilicate glass have a much lower mobility than alkali ions so that the charge relaxation in borosilicate glass is dominated by alkali ions \cite{MurayBC}. The charge relaxation current is thus due to impurity ions that are field driven away from the injected holes. The charge migration results in depleting or accumulating ions at the interfaces, without however accumulating charge in the bulk. } The electric field is thus divergence free in the bulk and sourced by the charged interface.

The dynamics of the accumulated charge is described by a surface charge dynamics at the inner and outer interfaces \cite{Giglio_PRA_2018,Giglio_PRA_2020}. The accumulated charges are field-driven along the interfaces as well as from the inner surface to the outer surface through the bulk. The grounded electrodes that cover the outer surface of the capillary  (see Fig. \ref{fig_profile}) act as absorbing boundary conditions for the deposited charge, allowing the deposited charge to be depleted. 
We simply note here that the surface charge dynamics in  \textsf{InCa4D}  depend  on three free parameters, the bulk conductivity $\kappa_b$, the surface conductivity $\kappa_s$  and dielectric constant $\varepsilon_r$ of the insulating capillary,  which need to be provided. 
Details about the theoretical model, on which the numerical code is based, can be found in \cite{Giglio_PRA_2018,Giglio_PRA_2019,Giglio_PRA_2020}. 

In order to stick to the experimental data presented in \cite{Giglio_PRA_2018}, we consider  a 2.3 keV Ar$^+$ beam and a borosilicate glass capillary, which is characterized at room temperature by a surface conductivity of $\kappa_s=10^{-16}$ S, a bulk conductivity of $\kappa_b=10^{-13}$ S/m and dielectric constant of $\varepsilon_r=4.6$ \cite{Gruber12}.  
In the experimental setup in \cite{Giglio_PRA_2018}, the capillary was surrounded by a grounded cylinder electrode of 6 mm  inner radius. The latter screened the capillary from stray electrons that would otherwise be attracted by the charged capillary and hinder the rise of the potential in the capillary \cite{Giglio_PRA_2017}. The grounded cylinder also imposed axisymmetric boundary conditions to the electric field, which is crucial for radial focusing.  For all these reasons, the capillary in our simulations is also surrounded by a grounded conducting cylinder, having an inner radius of 6 mm. Simulating the discharge of a previously charged capillary, we found that the capillary in configuration (INS) has a simulated charge depletion time of about 9 hrs, close to the 11 hrs found for the capillary in \cite{Giglio_PRA_2018}. 

Finally, the injected ion beam is supposed spatially uniform and characterized by its current intensity $I_\text{in}$ and root-mean-square (rms) emittance $\epsilon_\text{rms}$ \cite{Emittance}.  In the following, we will see how these two beam parameters influence the  fraction $I_\text{out}/I_\text{in} $ of the transmitted beam. In view of the beam intensities considered here, space charge forces are ignored. {\color{blue} Indeed, for injected intensities below $I_\text{in}<2$ pA, the  average distance separating two (2.3 keV) Ar$^+$ ions in the beam is more than 8 mm. We estimated that for such a low beam density, the trajectories can be considered independent from each other.}

\section{Self-organized radial focusing}

\subsection{Configuration (INS)}
\label{SEC_INS}

\begin{figure}[ht!]
\begin{center}
\includegraphics[scale=0.1]{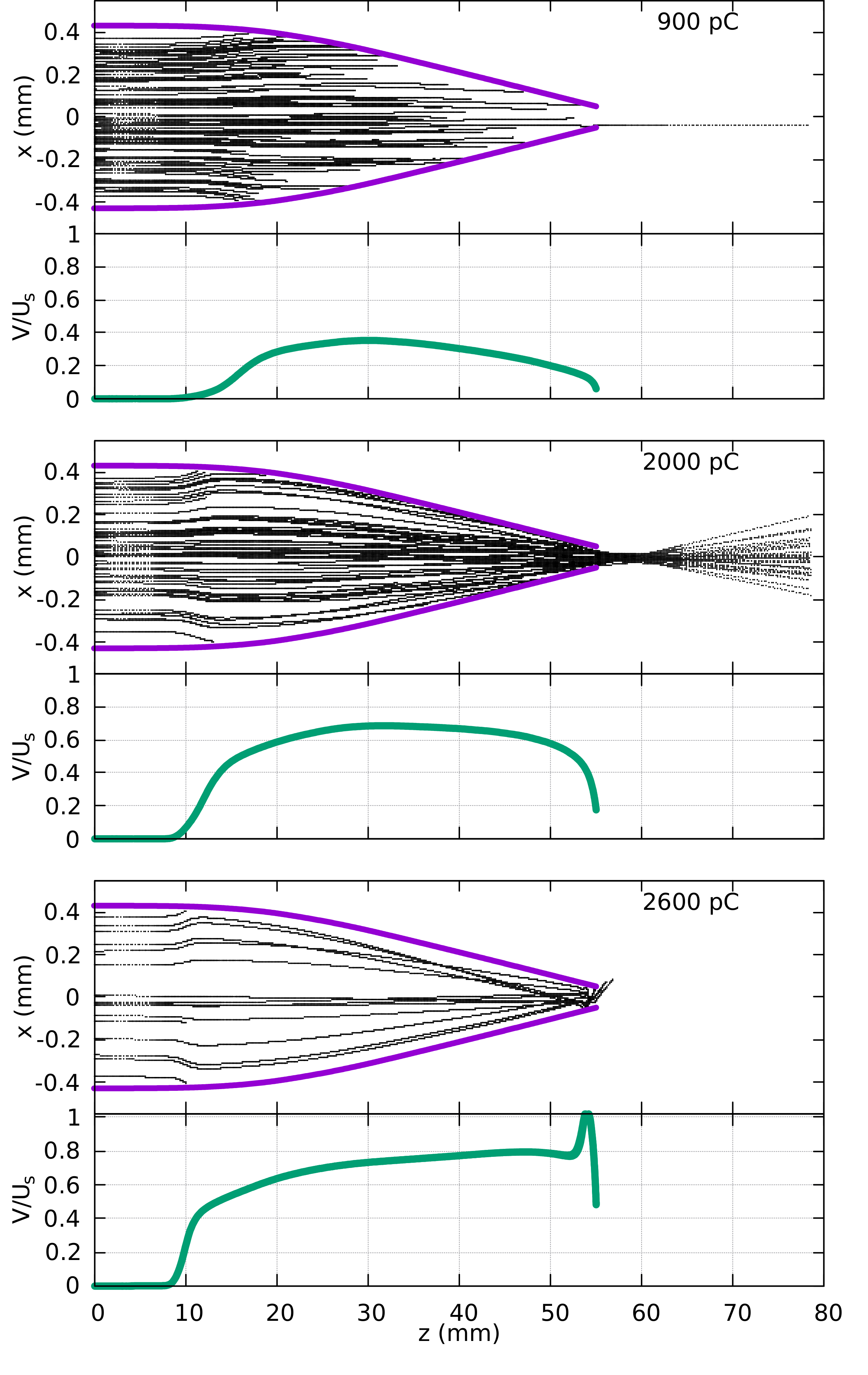}
\end{center}
\caption{Ion trajectories (black) at 3 different moments $Q/I_\text{in}$, identified by the injected charge $Q$. Beam intensity is $I_\text{in}=1$ pA.   Configuration (INS) was used and the profile of the inner surface of the capillary are shown in violet. Below each panel showing the trajectories, the electric potential, normalized with respect to the potential of the ion source $U_s$,  is given.}
\label{fig_focus_INS}
\end{figure}

We  briefly illustrate the self-organized radial focusing of an ion beam by a tapered glass capillary and discuss the Coulomb blocking that usually appears  with configuration (INS). We consider an injected beam current of $I_\text{in}=1$ pA with a rms emittance of  $\epsilon_\text{rms}=0.36$ mm.mrad (see definition Eq. \ref{eq_rms}). 
{\color{blue} The 3 panels of Fig. \ref{fig_focus_INS} show the  ion trajectories at different moments $Q/I_\text{in}$, indexed by the amount of injected charge $Q$. } The profile of the inner surface of the capillary is given by the violet curve. The green curve gives the electric potential in the capillary, taken at the inner surface, as a function of $z$.  {\color{blue}
As the first 8 mm of the outer surface of the capillary are grounded, and because  little to no charge is deposited by the beam in that area, the potential inside the capillary stays close to zero for $z<8$ mm. }

Initially, the transmitted fraction is about 1.3 \%, which is slightly lower than the geometrical fraction of $R_h^2/R_b^2 \simeq 1.6$ \% because of the non-zero beam divergence of the injected beam. After that $Q=900$ pC have been injected (first panel of Fig. \ref{fig_focus_INS}),  the transmitted fraction is still close to the geometrical one, and has only increased to 2.4 \%. The potential in the capillary reaches 900 V  at its highest point (40\% of the ion source potential $U_s=2300$ V), but is still too low to induce a non-negligible radial focusing. 

Once 2000 pC have been injected (middle panel of Fig. \ref{fig_focus_INS}), the potential in the capillary reaches in the region around $z=30$ mm a value of about 70\% of the extraction potential $U_s$. The self-organized potential is now strong enough to focus the injected beam, with the focusing point lying just behind the capillary outlet. Note the typical profile of the potential, which has a strong negative curvature (responsible for the focusing \cite{Giglio_PRA_2018}) and is close to zero at the inlet and outlet of the capillary, similar to an electrostatic Einzel lens \cite{Einzel}. At this point, the transmitted fraction is maximal and around 87\%, without however attaining 100\%. 

The part of the beam that is not transmitted will however continue to charge the capillary. We found that at the moment the transmission rate becomes maximal, the capillary is characterized by a leakage current of about $I_\text{leak} =70$ fA. As the deposited charge per unit time $I_\text{out}-I_\text{in} \simeq 130$ fA exceeds the leakage current, the beam will continue to charge the capillary. As a result, the electric potential over-focuses the injected beam and the ions hit now the inner surface  near the tip. The latter accumulates charge at higher rate and after 2600 pC have been injected (last panel of Fig.  \ref{fig_focus_INS}), the potential $V$ near the tip exceeds $U_s$ and the beam is Coulomb blocked. In configuration (INS) the Coulomb blocking is inevitable as long as the depletion (leakage)  current is dominated by the charge accumulation rate.

\subsection{Configuration (GND)}
\label{SEC_GND}

In the previous section, we identified that the Coulomb blocking of the transmitted focused beam  is due to a rapid charge accumulation near the tip. In order to avoid the Coulomb blocking, we propose to ground the tip by covering the outer surface of the tip with a conducting paint and connect it electrically to the ground. This yields configuration (GND), which we will consider in the following investigation.

\begin{figure}[ht!]
\begin{center}
\includegraphics[scale=0.44]{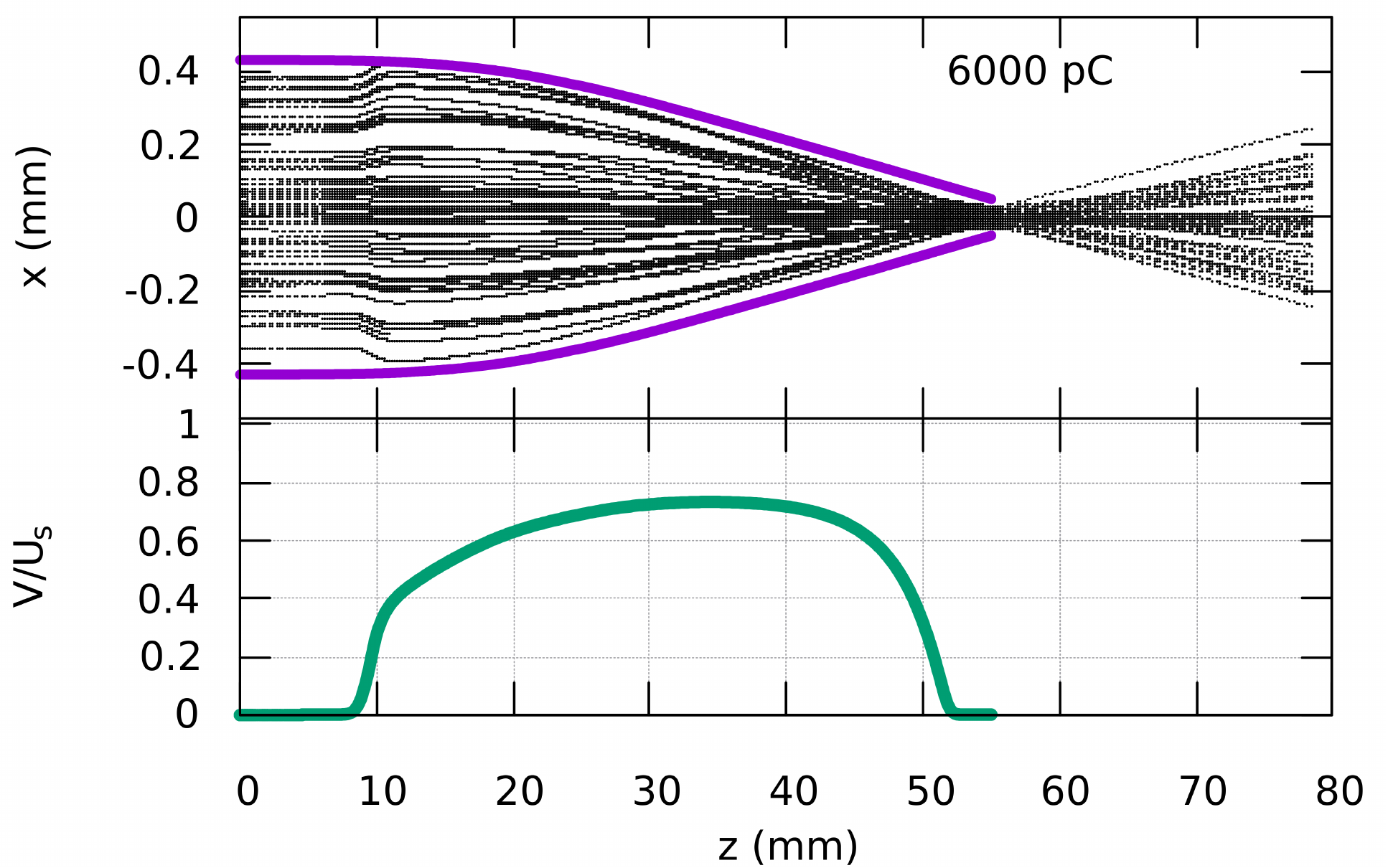}
\end{center}
\caption{Ion trajectories (black) after 6000 pC have been injected. Configuration (GND) was used and the profile of the inner surface of the capillary are shown in violet. Below we show the electric potential normalized with respect to the potential of the ion source $U_s$}
\label{fig_focus_GND}
\end{figure}

We consider once more a beam current of 1 pA with a rms emittance of  $\epsilon_\text{rms}=0.36$ mm.mrad and simulate the ion trajectories for the capillary in configuration (GND).  Again, we found that, once   $Q=2000$ pC has been injected, the beam is focused through the capillary, with a maximal transmitted fraction of 90\% and a transmitted current density compressed by a factor $f_c=57$!
However, and in contrast to configuration (INS), we do not experience Coulomb blocking, even after 6000 pC has been injected, see Fig.  \ref{fig_focus_GND}. Indeed, the grounded electrode that covers the outer tip of the capillary keeps the potential at the outlet close to zero. Charges that are injected by ions at the tip are efficiently screened by the electrons of the additional grounded electrode, avoiding the rise of the potential, unlike case (INS). For the time of the simulation ($Q>12000$ pC or equivalently $t>6$ hrs), the profile of the potential in the capillary seemed stationary. In particular, the location of the waist of the focusing "point" was kept at the capillary outlet. The self-organized charged capillary behaves thus like an electrostatic Einzel lens \cite{Einzel} plus a collimator hole placed at the focusing point. 

We expect thus the (GND) setup to stabilize the transmission of the focused beam, making  capillaries with such an add-on suitable to be used as self-organized electrostatic lenses. Focusing capillaries have the advantage over standard Einzel lenses of being quite compact while yielding little spherical aberration because the potential in the capillary is optimized to maximize the transmitted fraction. Also, unlike Einzel lenses, the potential needed to focus the beam at the outlet of the lens is well below the extraction potential of the source.  However, as we will see in the next section, the ability of our capillary to focus a beam  depends strongly on the beam parameters.

%

\section{Influence of the beam parameters}

\subsection{Beam intensity}
\label{SEC_INTENS}
It was already shown in \cite{Giglio_PRA_2018}, which uses a capillary in a configuration similar to (INS), that the intensity of the injected beam has a strong influence on the  transmission rate and on the moment the Coulomb blocking sets in. 
We will now check in how far configuration (GND) improves over (INS).  
Incidentally, we will also check if a conically shaped tip is preferable to the slightly funnel-shape one used in \cite{Giglio_PRA_2018}.

\begin{figure}[t!]
\begin{center}
\includegraphics[width=8.4cm]{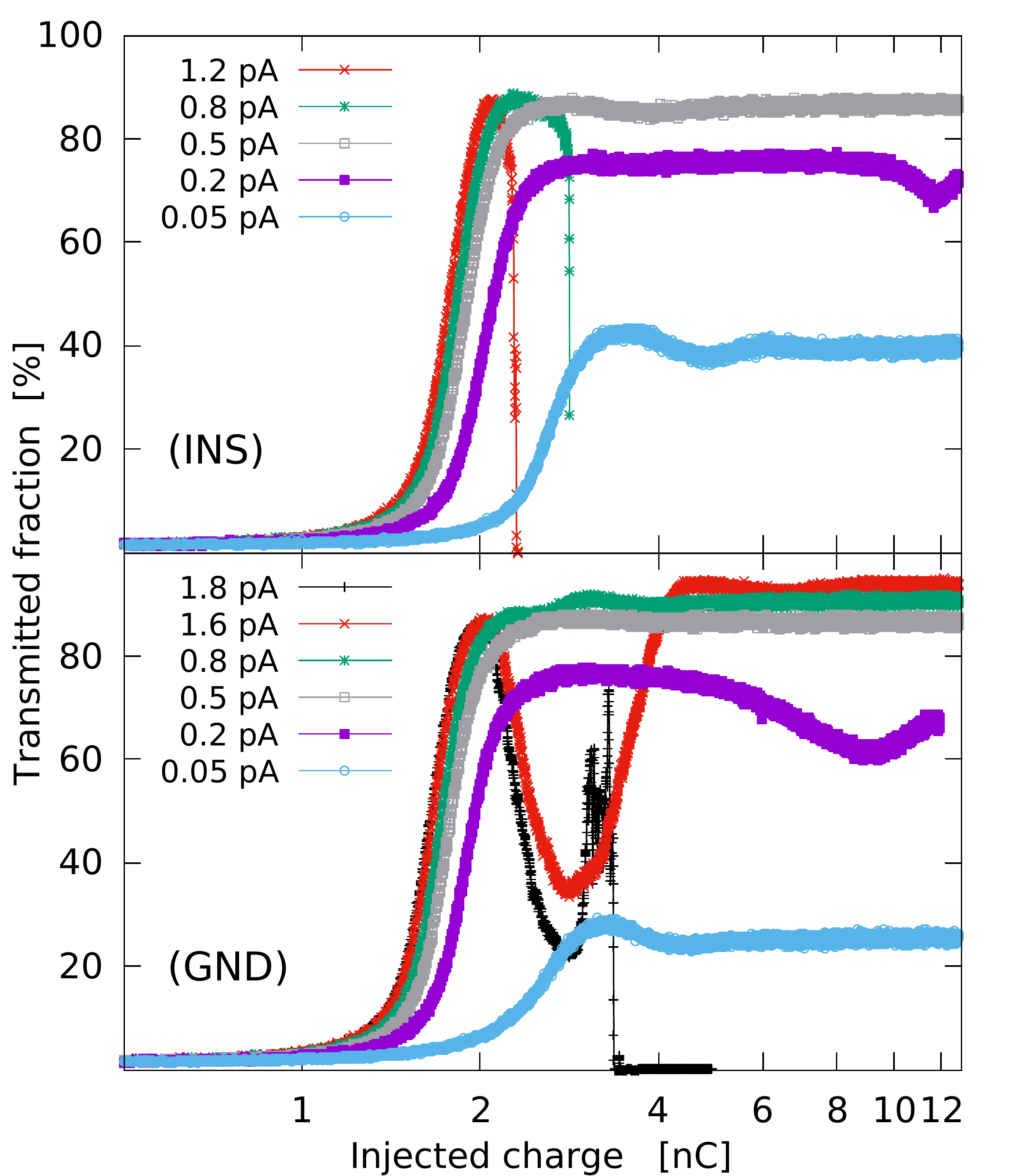}
\end{center}
\caption{Evolution of the transmitted fraction as a function of the injected charge for various beam intensities $I_\text{in} \in [0.05, 1.8]$ pA. Injected beam emittance was $\epsilon_\text{rms}=0.36$ mm.mrad. Upper panel used configuration (INS) and lower panel  configuration (GND). Note the x-log scale.} 
\label{fig_trans_intensity_INS_GND}
\end{figure}
In Figure \ref{fig_trans_intensity_INS_GND} we simulated for different intensities the transmitted fraction as a function of time or alternatively as a function of the injected charge $Q=I_\text{in}\times t$.
The injected current intensities range from 50 fA to 1.8 pA. Intensities lower than 50 fA are not relevant given the leakage current of about 70 fA found for setup (INS) at the moment the transmitted fraction is maximal.  The emittance of the injected beam was sufficiently small, $\epsilon_\text{rms}=0.36$ mm.mrad, as not to influence the observed trends, cf. subsection \ref{SEC_EMIT}.  
In the upper panel of Fig.  \ref{fig_trans_intensity_INS_GND}, the capillary was used in configuration (INS), while in the lower panel, it was used in configuration (GND).  

We note that, for (INS), the transmitted fraction increases with the beam intensity, reaching a maximum value of about 87 \%, or alternatively, a notable compression factor of $f_c=56$.
This trend was already explained in \cite{Giglio_PRA_2018} by the higher negative curvature of the electric potential generated by the deposited charge in the case of higher injected intensities. 
For $I_\text{in}=50$ fA, which deposes charge in the capillary wall at a rate well below the leakage current, the Coulomb blocking is easily avoided, but the asymptotic transmitted fraction with 40 \% is below optimum. 
For $I_\text{in} = 0.5$ pA, the transmitted fraction tops at 85 \% and the Coulomb blocking could be delayed beyond the simulated 12 nC of injected charge. Indeed, asymptotically, the deposited charge per unit time $I_\text{in}-I_\text{out}\simeq 75$ fA is balanced by the leakage current of about 70 fA. 
However, for even higher intensities, $I_\text{in}\ge 0.8$ pA, the Coulomb blocking sets in quickly after the maximum transmission rate was obtained. Simulations yield that for setup (INS) there is a narrow intensity window $I_\text{in} \in [0.2, 0.6]$ pA for which the focusing is near optimal and Coulomb blocking possibly avoided. 
Note that in \cite{Giglio_PRA_2018} no such intensity range window, yielding a stable transmission, was found. We attribute the lack of such "stability" window to the slightly funnel-shaped tip profile of the capillary used in \cite{Giglio_PRA_2018}. This indicates that a conical tip profile is to be favored, which is not surprising seeing that the trajectories are focused by the field in the first 30 mm of the capillary and then are almost in free flight until they reach the outlet.

For setup (GND), we found that beam intensities up to 1.6 pA are no longer Coulomb blocked, meaning that the grounded tip allowed  enlarging the relevant "stability" window, $[0.2, 1.6]$ pA, by a factor 3.5 with respect to setup (INS).  For intensities above 1.2 pA,  we notice however a drop in the transmitted fraction, just after 2 nC of injected charge, as can be seen for $I_\text{in} =1.6$ (red curve) in the bottom panel of Fig. \ref{fig_trans_intensity_INS_GND}. The drop in the transmission is due to over-focusing. But because of the grounded tip, the Coulomb blocking is avoided and the charge distribution has sufficient time to re-organize itself and stabilize the transmission, which asymptotically peaks at 92 \%. The current intensity  window that yields a stable transmission of a focused beam was thus significantly enlarged by grounding the tip of the capillary. The addition of the grounded tip electrode is thus a step in the right direction to  allow  tapered capillaries to be used to focus ion beams in the keV range.  We note that the intensity window $[0.2,1.6]$ pA, yielding high and stable transmission fractions, can be shifted towards higher intensities, simply by using a higher conducting glass capillary. 

Interestingly, for $ I_\text{in}  \ge 1.8 $ pA, the transmission rate becomes erratic after the maximum transmission is obtained, (around 2 nC ) and the Coulomb blocking sets in after 3 nC have been injected. The reason is that the induced electric potential over-focuses the beam and the focusing point moves beyond the 3 mm grounded tip. As a result, beam  ions hit and deposit charge in the part of the capillary that is not covered by the 3 mm of conducting paint. We expect thus that the length of the grounded tip has an influence on the highest possible intensity that can be stabilized.  

\begin{figure}[t!]
\begin{center}
\includegraphics[width=9cm]{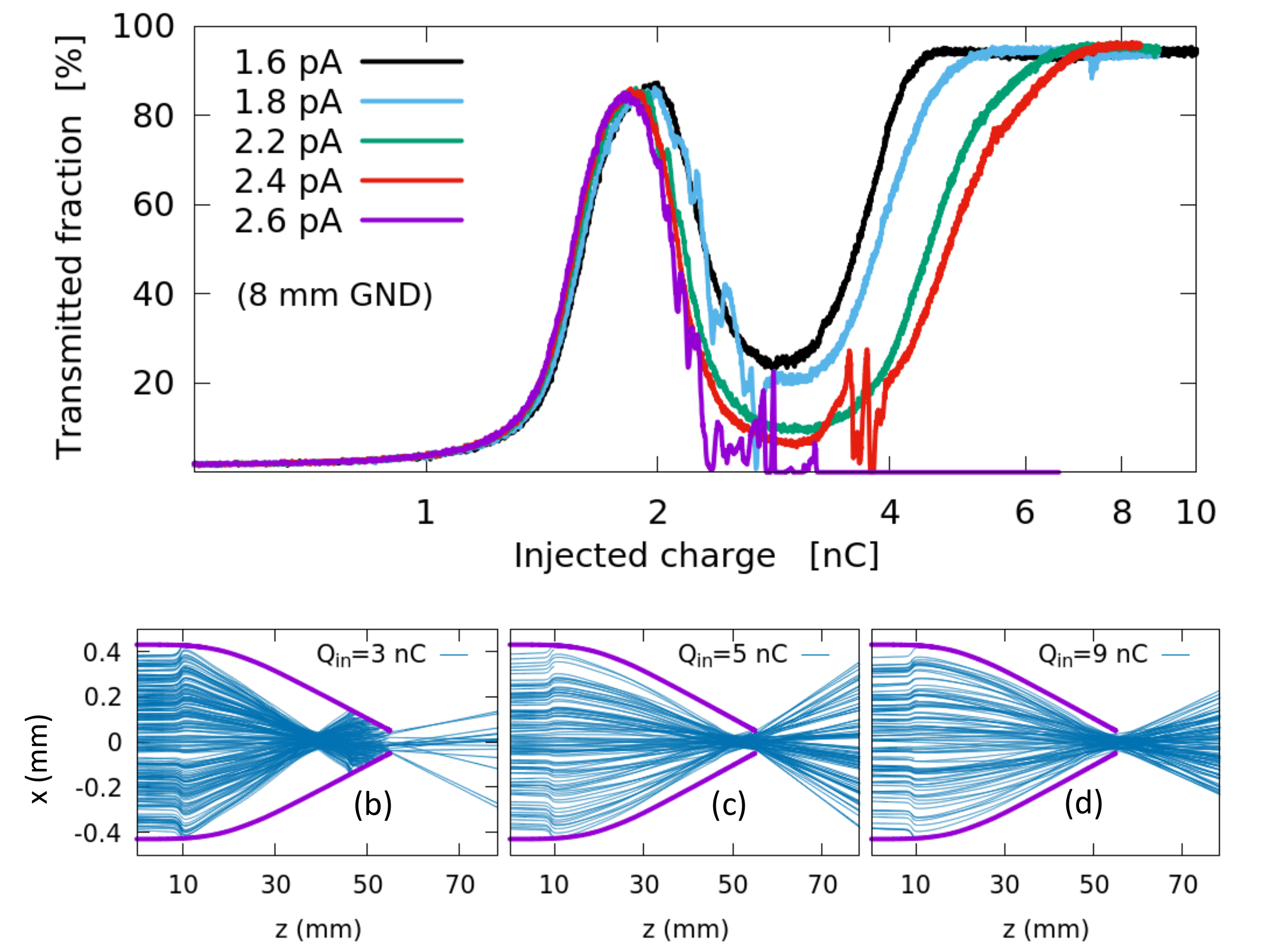}
\end{center}
\caption{Panel (a): evolution of the transmitted fraction as a function of the injected charge $Q$ for various beam intensities $I_\text{in} \in [1.6, 2.6]$ pA. Configuration (GND) is used but with a grounded tip electrode which is 8 mm long. Panels (b), (c) and (d) show the trajectories at different moments indicated by the injected charge $Q_\text{in}$, for the case $I_\text{in}=2.4$ pA.} 
\label{fig_long_GND}
\end{figure}

{\color{blue}
We checked the latter claim, by simulating the time evolution of the transmission rate  in the case where the last 8 mm of the tapered capillary were grounded. The results for $I_\text{in}  \in [1.6,2.6]$ pA are given in Fig. \ref{fig_long_GND}a. For this range of intensities, the deposited charges generate an electric field that tends to over-focus the beam so that a large fraction of the beam ions are intercepted by the tip (Fig. \ref{fig_long_GND}b). For intensities  $I_\text{in}\le 2.4$ pA, the 8 mm long grounded electrode covering the outer surface of the tip avoids that  the deposited charges 
generate a potential high enough to Coulomb block the beam. This amounts to sufficient time for the charge dynamics to self-organize the surface charge distribution, so that the focusing point moves back toward the capillary exit, maximizing and stabilizing the transmission (Fig. \ref{fig_long_GND}c - \ref{fig_long_GND}d). However, for $I_\text{in} \ge 2.6$ ( violet curve of Fig. \ref{fig_long_GND}a ), the over-focused beam deposits charge in a region of the tip that is not covered by the 8 mm electrode, resulting in a Coulomb blocking of the beam.  We conclude that a longer grounded tip-electrode allows indeed higher beam intensities to be transmitted. In would be interesting in the future to check if an optimal value exists.}

\subsection{Beam emittance}
\label{SEC_EMIT}

Apart from the current intensity, the emittance of the injected beam was also found to have a strong influence on the transmission rate. To the authors' knowledge, the influence of the emittance on the guiding power in insulating capillaries was never studied before, mainly because in straight capillaries the divergence rather than the emittance of the beam is considered relevant. As the emittance controls the minimum radius of the waist of the focusing point, it is easily seen that in order to have an optimal transmission rate, the emittance of the beam must be lower than a threshold value determined by the outlet radius $R_h$ of the tapered capillary.
\begin{figure}[h!]
\begin{center}
\includegraphics[width=8.0cm]{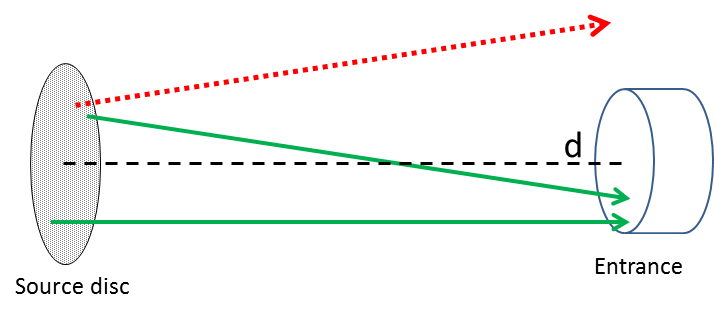}
\end{center}
\caption{Modeling of the beam emittance. Positions of  projectiles are uniformly sampled on the source disc and emitted  at a given angle, normally distributed. Trajectories that miss the entrance (red doted arrow) are discarded. The green arrows are the trajectories that enter the capillary. }
\label{fig_emittance}
\end{figure}

We simulated the transmitted fraction as a function of the injected charge for various beam emittances, expressed in mm.mrad. To avoid ambiguities, we define the geometrical root mean square emittance $\epsilon_\text{rms}$  used in this work,
\begin{equation}
\epsilon_{\text{rms}}=4 \sqrt{<x x> <x' x' > - <x x' > ^2} \quad,
\label{eq_rms}
\end{equation}
with
\begin{equation}
<x x' > =\overline{(x-\overline{x})(x'-\overline{x'})} \quad,
\label{eq_var}
\end{equation}
where  $x'=v_x/v_z$ and $v_z$ the velocity along the beam axis \cite{Emittance}. The overline operator stands for the mean value of the distribution.  The initial phase space distribution of the injected beam were selected according to the scheme in Fig. \ref{fig_emittance}. The initial positions of the ions were sampled uniformly on a disc having a diameter of 2 mm and located  a distance $d$ upstream from the capillary entrance. The angle of the velocity vectors were sampled according to a normal distribution having a FWHM of 0.3 deg. The projectiles that, after traveling through the field free region of distance $d$, miss the entrance of the capillary (red doted arrow), are discarded. Only the trajectories that enter the capillary (green arrows) are retained in the calculation of the rms emittance.
The beam emittance is then varied simply by modifying the distance $d$ between the source and the entrance of the capillary. In this work, the geometrical rms emittance $\epsilon_\text{rms}$ of the injected beam was varied from 1.56 to 0.36 mm.mrad by changing the distance $d$ from 20 to 100 cm.

\begin{figure}[h!]
\begin{center}
\includegraphics[width=8.4cm]{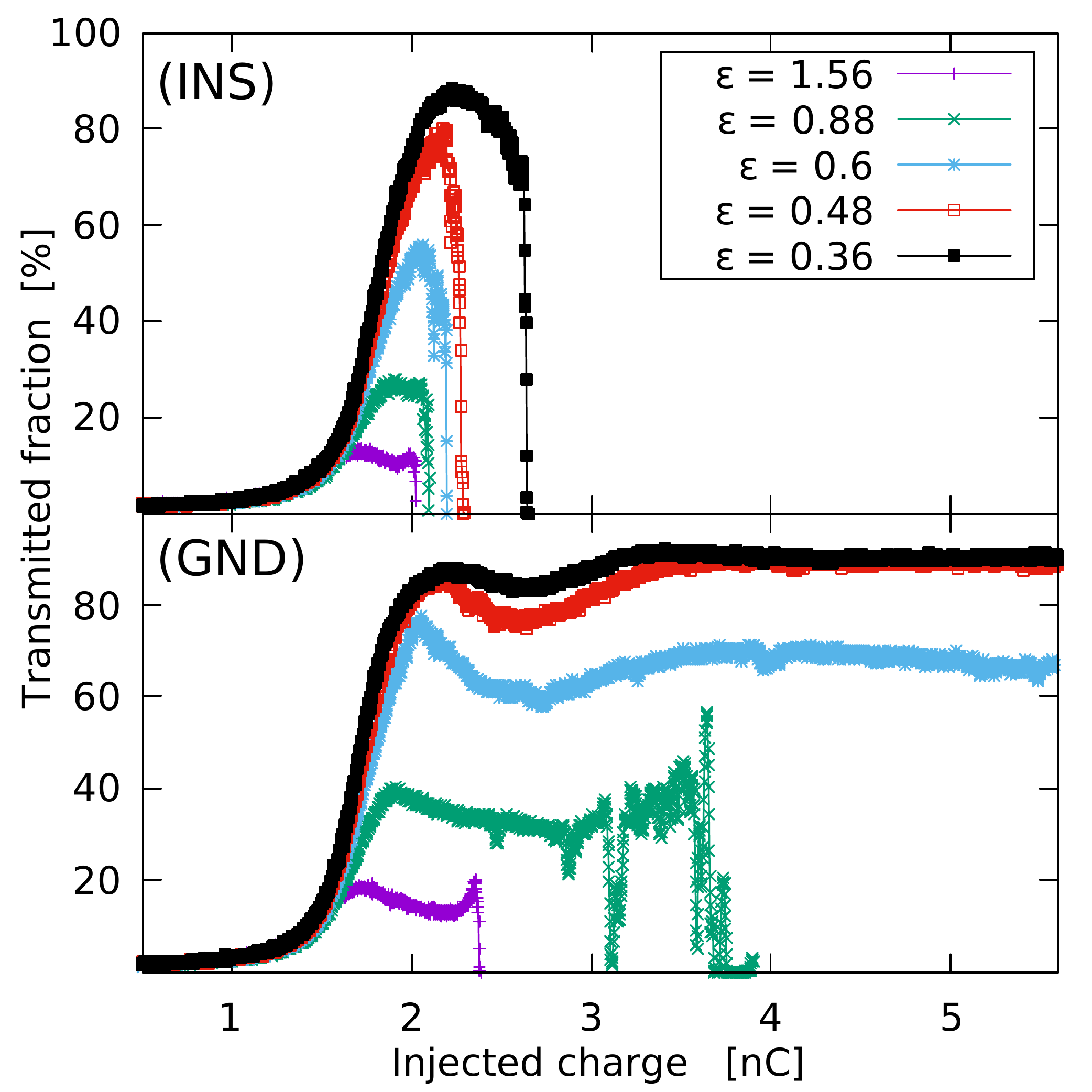}
\end{center}
\caption{Evolution of the transmitted fraction as a function of the injected charge for various emittances $\epsilon \equiv \epsilon_\text{rms} \in [0.3,1.6]$ mm.mrad. The chosen configuration is shown in the top left corner.  Intensity of the injected current was 1 pA.} 
\label{fig_trans_elittance_INS_GND}
\end{figure}

In the following simulations, the beam intensity was set to $I_\text{in}=1$ pA, for which over-focusing is expected.  In the upper panel of Fig. \ref{fig_trans_elittance_INS_GND}, we record the transmitted fraction as a function of the emittance for a capillary in configuration (INS). 
Without surprise, we note that Coulomb blocking appears whatever the value of the beam emittance. The main result here is that the peak value of the transmitted fraction decreases with increasing emittance. This is easily understood by the following interpretation: let us 
attribute an "acceptance" to the capillary of $a=0.54$ mm.mrad, corresponding to the outlet radius $R_h=0.05$ mm times the half-opening angle $\beta=10.8$ mrad of the tip.
If the injected beam emittance exceeds the acceptance of the capillary, $\epsilon_\text{rms}>a$, we expect that the radius of the waist of the focusing point exceeds the outlet radius and a part of the beam is inevitably intercepted by the tip, reducing thus the maximal value of the transmitted fraction. 

In the lower panel of Fig. \ref{fig_trans_elittance_INS_GND}, the transmitted fraction is recorded for setup (GND). 
For beams having $  \epsilon_\text{rms} \le 0.6$ mm.mrad, the grounded tip avoids Coulomb blocking. In particular, one notices that for  $\epsilon_\text{rms}=$ 0.36 (black curve) and 0.48 mm.mrad (red curve) the maximal transmitted fraction of 90\% is reached, confirming that the waist of the focusing point for $\epsilon_\text{rms} < a$ fits within the  50 $\mu$m of the outlet diameter of the capillary. However, for  $\epsilon_\text{rms} \ge 0.8>a$, the size of the waist is such that ions hit also the part of the capillary that is not covered by the tip electrode. As a result, charge accumulates rapidly in the uncovered region of  the tip, generating locally a potential that exceeds the source potential $U_s$.  This seems to indicate that if a larger part of the tip was grounded, a beam with an emittance of  $\epsilon_\text{rms} = 0.88 $ (green curve) could be stabilized.

\section{Conclusion}

We proposed a numerical study on the fraction and stability of the transmitted beam after self-organized focusing sets in, due to the accumulated charge in the capillary.  
We considered two configurations for our conically tapered capillary, which differed by the 3 mm long grounded electrode that covers the capillary tip. We briefly summarized the self-organized radial focusing mechanism and  the origin of the Coulomb blocking in tapered capillaries. Then, we simulated the transmitted fraction as a function of the injected charge for different beam intensities and beam emittances. For configuration (INS), we identified a narrow range of beam intensities and emittances that yields transmitted fractions (up to 85 \%)  and avoids Coulomb blocking. We attribute the existence of such a "stability" window to the conical-shaped capillary tip. 
The narrow intensity range however suggests that a balance between the deposited and leakage current is difficult to maintain.

We showed that setup (GND) is an improvement over setup (INS), as it avoids the Coulomb blocking for a larger range of intensities and emittences. In the particular case, where the beam emittance is lower than the capillary acceptance, a transmitted fraction of 92 \% could be stabilized for an injected beam intensity of $I_\text{in}=1.6$ pA, which was not possible in setup (INS).  Remarkably, with the given dimensions of the capillary, a compression factor up to $f_c=56$ was found. 

The simulated results also indicate that the length of the grounded electrode that covers the tip has an influence on the range of beam intensities and emittances that can be stabilized. {\color{blue} In the case of a 8 mm long grounded tip electrode, we found that the Coulomb blocking could be avoided for injected beam intensities  up to 2.4 pA, which is already a significant improvement over the 1.6 pA, obtained in the case of the 3 mm short grounded tip electrode. }
It would thus be interesting to check what the optimal length of the tip electrode is. Our work suggests that grounding the tip of the capillary may be part of the solution for realizing an electrostatic self-organized lens for the generation of  stable focused micro-beams. 
%
%
Finally, we note that the presented numerical predictions can be tested and validated experimentally in a dedicated experiment. Work in this direction is underway.

\begin{acknowledgments}
This work was supported by the French research agency
Centre National de la Recherche Scientifique via Project
International de Coopération Scientifique Project No. 245358
Hongrie 2018.
\end{acknowledgments}

\end{document}